\documentclass[conference]{IEEEtran}

\ifCLASSINFOpdf
\else 
\fi

\usepackage{url}
\usepackage{cite}
\usepackage{amsmath,amssymb,amsfonts}
\usepackage{algorithmic}
\usepackage{graphicx}
\usepackage{textcomp}
\usepackage{booktabs}
\usepackage{multirow}
\usepackage{xcolor}
\usepackage[]{fancyhdr} %
\newcommand{\changefont}{\fontsize{9}{9}\selectfont}
\IEEEoverridecommandlockouts
\begin{document}
	\bstctlcite{IEEEexample:BSTcontrol}
	\bibliographystyle{IEEEtran}
%
\title{Deep Reinforcement Learning for Long-Term Voltage Stability Control}

\author{\IEEEauthorblockN{Hannes Hagmar}
\IEEEauthorblockA{\textit{Department of Electrical Engineering} \\
\textit{Chalmers University of Technology}\\
Gothenburg, Sweden \\
hannes.hagmar@chalmers.se}
\and
\IEEEauthorblockN{Le Anh Tuan}
\IEEEauthorblockA{\textit{Department of Electrical Engineering} \\
\textit{Chalmers University of Technology}\\
Gothenburg, Sweden \\
tuan.le@chalmers.se}
\and
\IEEEauthorblockN{Robert Eriksson}
\IEEEauthorblockA{\textit{Department of Power Systems} \\
\textit{Swedish National Grid}\\
Sundbyberg, Sweden \\
robert.eriksson@svk.se}
}


%





\maketitle
\thispagestyle{fancy}
\pagestyle{fancy}

\begin{abstract}

Deep reinforcement learning (DRL) is a machine learning-based method suited for complex and high-dimensional control problems. In this study, a real-time control system based on DRL is developed for long-term voltage stability events. The possibility of using system services from demand response (DR) and energy storage systems (ESS) as control measures to stabilize the system is investigated. The performance of the DRL control is evaluated on a modified Nordic32 test system. The results show that the DRL control quickly learns an effective control policy that can handle the uncertainty involved when using DR and ESS. The DRL control is compared to a rule-based load shedding scheme and the DRL control is shown to stabilize the system both significantly faster and with lesser load curtailment. Finally, when testing and evaluating the performance on load and disturbance scenarios that were not included in the training data, the robustness and generalization capability of the control were shown to be effective. 
\end{abstract}

\begin{IEEEkeywords}
Deep reinforcement learning, emergency control, voltage stability, optimal control, real-time control
\end{IEEEkeywords}

%
\IEEEpeerreviewmaketitle

\section{Introduction}

Long-term voltage instability has caused several major blackouts in the past and is a major aspect of power system security assessment \cite{Cutsem1998}. In the case of a larger disturbance, the ability of system operators to act quickly and with the correct control measures is imperative to avoid a fully developed voltage collapse. Voltage stability control typically includes actions such as generation redispatch or tripping, capacitor/reactor switching, excitation boosting, load shedding, or controlled system separation \cite{Kundur1997}. Choosing efficient and suitable control actions, which can both mitigate instability \textit{and} minimize the impact on the end-consumers, can significantly improve the operational efficiency during adverse events. 

While some actions are automatically triggered by local protection schemes, some are required to be manually initiated by the system operators. Those protection actions are usually based on fixed settings that are pre-determined through off-line simulations of anticipated contingency scenarios and forecasted system conditions \cite{Huang2020}. Long-term voltage stability (LTVS) events are often deceiving and the system may seem stable only to end up in an unstable state within a short time \cite{Glavic2011}. Thus, once instability has been detected, the remaining time to evaluate the system condition and to choose suitable control actions is limited and can be overwhelming for system operators. The current system situation may also significantly differ from any of the previously studied off-line contingency scenarios and there is a risk that the actions that are taken are not sufficient in restoring the system's stability. 

Power system control is a problem of dynamic and sequential decision-making under uncertainty \cite{Huang2020}. Traditional methods based on optimal control (e.g., model predictive control), have difficulties in handling large dynamic models of real power systems. To be able to compute the optimal control actions in a time frame required by system operators, significant simplifications of the system model are then generally required. However, deep reinforcement learning (DRL) has in recent years shown significant progress in solving high-dimensional and complex control problems. It is based on having a control agent learn an optimal control policy through interactions with a real power system or its simulation model \cite{Ernst2004, Zhang2020}, where the combination with deep learning models allows it to handle large and continuous state spaces. Previous implementations of DRL in emergency control include methods for dynamic breaking \cite{Huang2020}, optimal load shedding for short-term voltage stability \cite{Zhang2018, Jiang2019, Huang2020}, automatic voltage control \cite{Diao2019,Wang2020,Thayer2020,Toubeau2020}, and oscillation damping \cite{Hashmy2020}. 

In this paper, we develop a DRL method for fast, optimal, and adaptive control for LTVS events. The method can in real-time suggest optimized control actions to system operators to stabilize the system. The DRL agent continuously monitors the system state, and if the taken actions are not sufficient, additional corrective actions are proposed. The main contributions are the following: 

\begin{itemize}
	\item A novel methodology for a DRL-based control for LTVS. The developed DRL agent can in real-time suggest optimized control actions to system operators to mitigate voltage instability. The problem formulation and the reward scheme are designed to incentivize a control that quickly stabilizes the system at a minimal cost. 
	
	\item An evaluation of using system services from demand response (DR) and energy storage systems (ESS) as a more economic and flexible alternative to stabilize the system. The paper specifically examines the capability of the DRL control to account for the uncertainty involved in using such services (e.g., the price and the availability) as an alternative to emergency load shedding. 
	
	\item An evaluation of the method's robustness and capability of handling scenarios that have not been included in the training data of the algorithm. This is important since the number of states and the possible combinations with different disturbance scenarios are very large for real power systems. 
\end{itemize}

The rest of the paper is organized as follows. In Section~\ref{sec:DRL}, the theory in DRL is presented. In Section \ref{sec:Method}, the proposed method is presented along with the steps for developing the training data and the training of the DRL algorithm. In Section \ref{sec:Simulations}, the results and discussion are presented. Concluding remarks are presented in Section \ref{sec:Conclusions}.

\section{Overview of Deep Reinforcement Learning}
\label{sec:DRL}

In DRL, a control agent uses its control policy to interact with an environment (or a system) to give a trajectory of states, actions, and rewards. The received reward - also commonly referred to as the reinforcement signal - is used to determine whether the taken actions were effective. Through continuous interactions with the environment, the agent is trained to maximize the expected sum of future rewards over time \cite{Silver2014}. 

In this study, we assume to have a stochastic policy $\pi(a|s)$ which models the conditional distribution for action $a \in \mathcal{A}$ given a state $s \in \mathcal{S}$. At each time step $t$, the agent observes the current state $s_t$ and samples an action $a_t$ from the policy. Once the action is taken, the environment responds with a reward $R_t$ and a new state $s' = s_{t+1}$ determined by a state transition dynamics distribution $p(s'|s,a)$. For a parametrized policy $\pi_\theta(a|s)$, the goal of the agent is to learn the optimal parameters $\theta^*$ that maximizes a defined objective function $J(\theta)$. The objective function is commonly defined to be \textit{expected} return, where the return, denoted $G^\gamma_t$, is the total discounted reward from a time step $t$ and onward:
\begin{equation} \label{eq:Return}
G_t^\gamma = \sum_{k=t}^T \gamma^{k-t}R_k
\end{equation}
and where $0<\gamma<1$ is a discounting factor. The value function is defined as the expected total discounted reward in state $s$ when following the policy:~$V^\pi(s) = \mathbb{E}\left[G_t^\gamma|s_t=s;\pi \right]$. The action-value function is defined as the expected total discounted reward in state $s$ when taking action $a$ and \textit{then} following the policy:~$Q^\pi(s,a) = \mathbb{E}\left[G_t^\gamma|s_t=s, a_t=a;\pi \right]$.

In this study, we will use the Proximal policy optimization (PPO) algorithm, first presented in \cite{Schulman2017}. The PPO algorithm has in previous studies been shown to provide good sample efficiency and learning stability while being capable of controlling both discrete and continuous actions variables depending on the problem formulation. We use the "clipped" version of the PPO algorithm, where the objective is defined as: 
\begin{equation} \label{eq:PPO_clipped}
\resizebox{.89\hsize}{!}{$J^{clip}(\theta)= \hat{\mathbb{E}}_t \left[\min\left(r_t(\theta) \hat{A}_t, \text{clip}\left(r_t(\theta), 1-\epsilon, 1+\epsilon\right) \hat{A}_t\right) \right]$}
\end{equation}
where $r_t$ is a probability ratio given by: 
\begin{equation} \label{eq:p_ratio}
r_t = \frac{\pi_\theta(a_t|s_t)}{\pi_{\theta_{\text{old}}}(a_t|s_t)}
\end{equation}
and $\epsilon$ governs the clipping range of the objective function, and $\theta_{\text{old}}$ refers to the vector of policy parameters used in sampling the transitions and thus before any update of the policy parameters. $\hat{A}_t$ is an estimator of the advantage function at time step $t$, given by
\begin{equation} 
\begin{split}
\label{eq:est_AdvFunc}
\hat{A}_t =  \hat{Q}_\phi^\pi(a_t, s_t)-\hat{V}_\phi^\pi(s_t) = \\ 
= \underbrace{R_t+\gamma\hat{V}_\phi^\pi(s_{t+1})-\hat{V}_\phi^\pi(s_t)}_\text{$\delta$-error}
\end{split}
\end{equation}
where $\hat{V}_\phi^\pi$ and $\hat{Q}_\phi^\pi$ are estimates of the value function and the action-value function, respectively. The action-value can be written recursively as the sum of the immediate reward $R_t$ after taking action $a_t$ in state $s_t$ and the estimated discounted value of the subsequent state $\gamma\hat{V}_\phi^\pi(s_{t+1})$. When this recursive expression is used to form the advantage, it is commonly referred to as the $\delta$-error (or temporal-difference error). 

The clipped objective function $J^{clip}(\theta)$ ensures that one does not move \textit{too} far away from the current policy, which allows one to run multiple epochs of gradient ascent on the samples without causing destructively large policy updates. The $r_t$-ratio is always equal to $1.0$ before the first epoch, when current policy $\pi_\theta(a_t|s_t)$ is the same as was used to sample the transitions $\pi_{\theta_{\text{old}}}(a_t|s_t)$. For each epoch, the policy is trained to \textit{increase} the probability ratio $r_t$ above 1.0 whenever the advantage function is \textit{positive}, thus making advantageous actions more probable to be chosen by the policy in the future. Similarly, the policy is trained to \textit{decrease} the probability ratio $r_t$ below 1.0 when the advantage function is \textit{negative}, thus making disadvantageous actions less probable to be chosen by the policy in the future. 

By computing the gradient of the objective function (typically by using automatic differentiation software such as Tensorflow or PyTorch), one can adjust the current policy through stochastic gradient ascent (or by alternatives such as Adam \cite{Kingma2014}) so that the defined objective function is maximized. The value function used to compute the advantage function in (\ref{eq:est_AdvFunc}) is generally unknown and has to be learned simultaneously as the policy. If the value function is learned in addition to the policy and the $\delta$-error is used to approximate the advantage function, the algorithm is usually referred to have an actor-critic architecture. The policy $\pi_\theta$ is estimated by the \textit{actor} while the value function $\hat{V}^\pi_\phi$ is estimated by the \textit{critic}. The role of the critic is thus to evaluate the effectiveness of the actions taken by the actor. The value function can be learned by forming and minimizing a new cost function, $L(\phi)$, based on the mean-squared error (or some other loss function) of the sampled and computed $\delta$-errors. In DRL, the capability of high dimensional feature extraction and non-linear approximation that deep learning and neural networks (NNs) provides is utilized. The parameters used in forming the policy ($\theta$) and the value function ($\phi$) are in this paper representing the node weights of two separate neural networks, and the goal of training the networks is thus to find the optimal node weights for these networks. 

\section{DRL for Long-term Voltage Stability Control}
\label{sec:Method}

The study period of interest in LTVS events may extend to several minutes and is typically driven by the actions of load tap changers, overexcitation limiters, and restorative loads \cite{Cutsem1998}. The relatively long time for instability to develop allows system operators to sample and process system measurements and to utilize centralized control strategies to achieve a more efficient control to mitigate instability. However, LTVS control in a large-scale power system is a highly non-linear and non-convex optimal decision-making problem. At every time instant, the controller (in DRL: the agent) should assess the state of the system and choose an action that can most efficiently stabilize it to the lowest system cost. The complexity lies both in interpreting the state of the system \textit{and} to determine what action is most optimal to take in that current state. The method of DRL control for LTVS is based on off-line training on a large data set consisting of dynamic simulations for a range of different disturbance and load scenarios. By training a DRL agent on those scenarios, a policy $\pi_\theta$ is developed which can form a mapping from the state (provided by measurements in the system) to an action. Once the DRL agent has been trained, it can in real-time suggest optimized control actions to system operators to mitigate voltage instability. 

\begin{figure}[t!]
	\begin{center}\vspace{-0.3cm}
		\includegraphics[width=7.8cm]{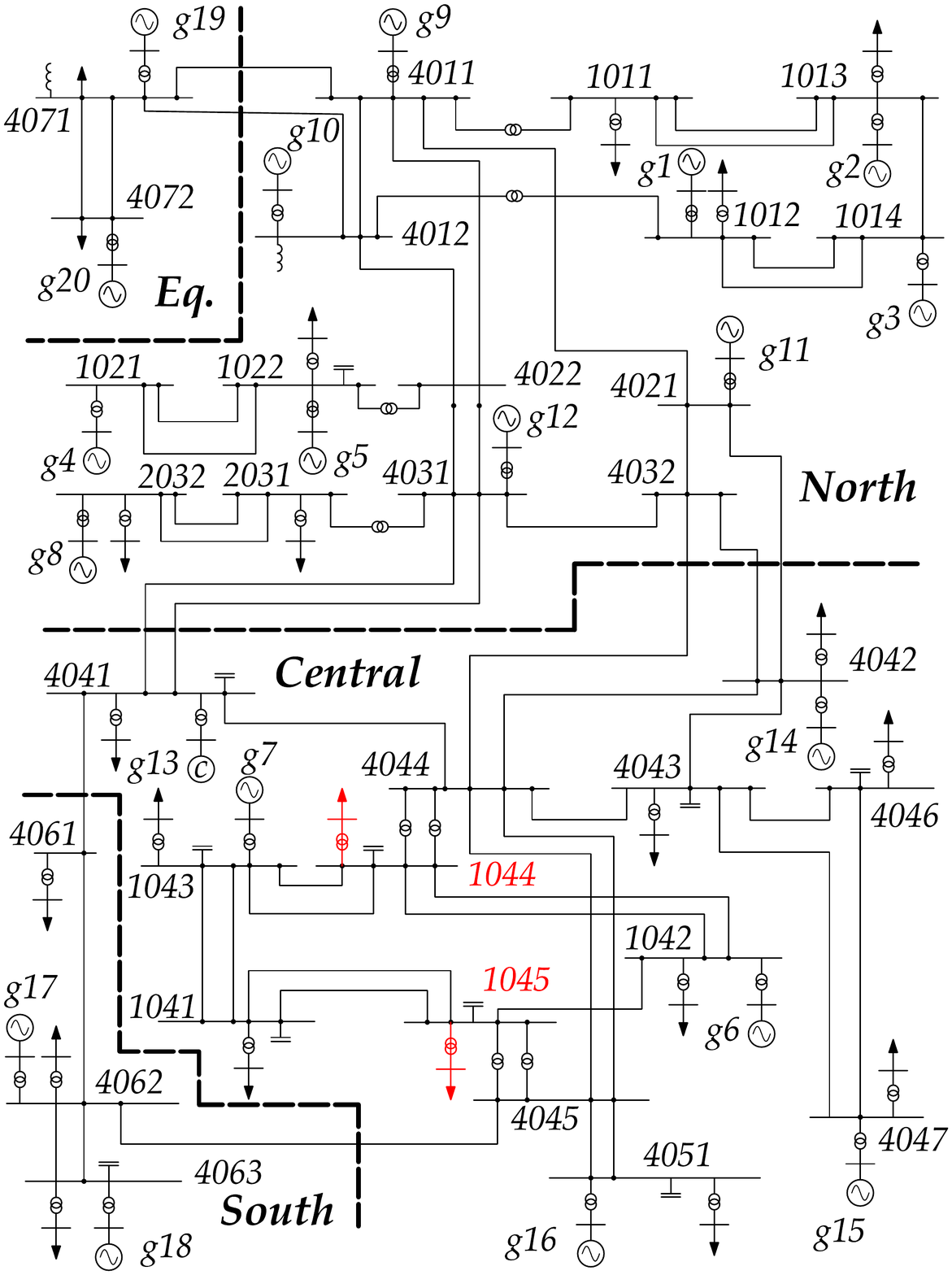}    
		\vspace{-0.6cm}
		\caption{One-line diagram of the modified Nordic32 system \cite{CutsemTestSystem2020}. Load buses 1044 and 1045 that participate in the load curtailment are marked in red.} 
		\label{fig:Nordic32_RL}
		\vspace{-0.4cm}
	\end{center}
\end{figure}

The training data are generated using PSS\textsuperscript \textregistered E 35.0.0 with its in-built dynamic models \cite{PSSEModel}. Alternatively, since the control is aimed at the LTVS phenomenon, a quasi-steady state (QSS) model could be used to generate the training data. A QSS model is sufficient to capture the essential behavior of a power system by replacing the short-term dynamics with their equilibrium equations and concentrating on long-term phenomena \cite{Cutsem1998}. All simulations have been tested on the modified version of the Nordic32 test system, detailed in \cite{CutsemTestSystem2020}. The test system is characterized by sensitivity towards long-term voltage instability. A one-line diagram of the test system is presented in Fig.~\ref{fig:Nordic32_RL}. In the following sections, all the details in generating the training data and the development of the DRL control are presented.

\subsection{Training data generation}
\label{sec:GenerationOCs}
\begin{figure}
	\begin{center}\vspace{-0.0cm}
		\includegraphics[trim=175 3 175 3,clip,width=8.6cm]{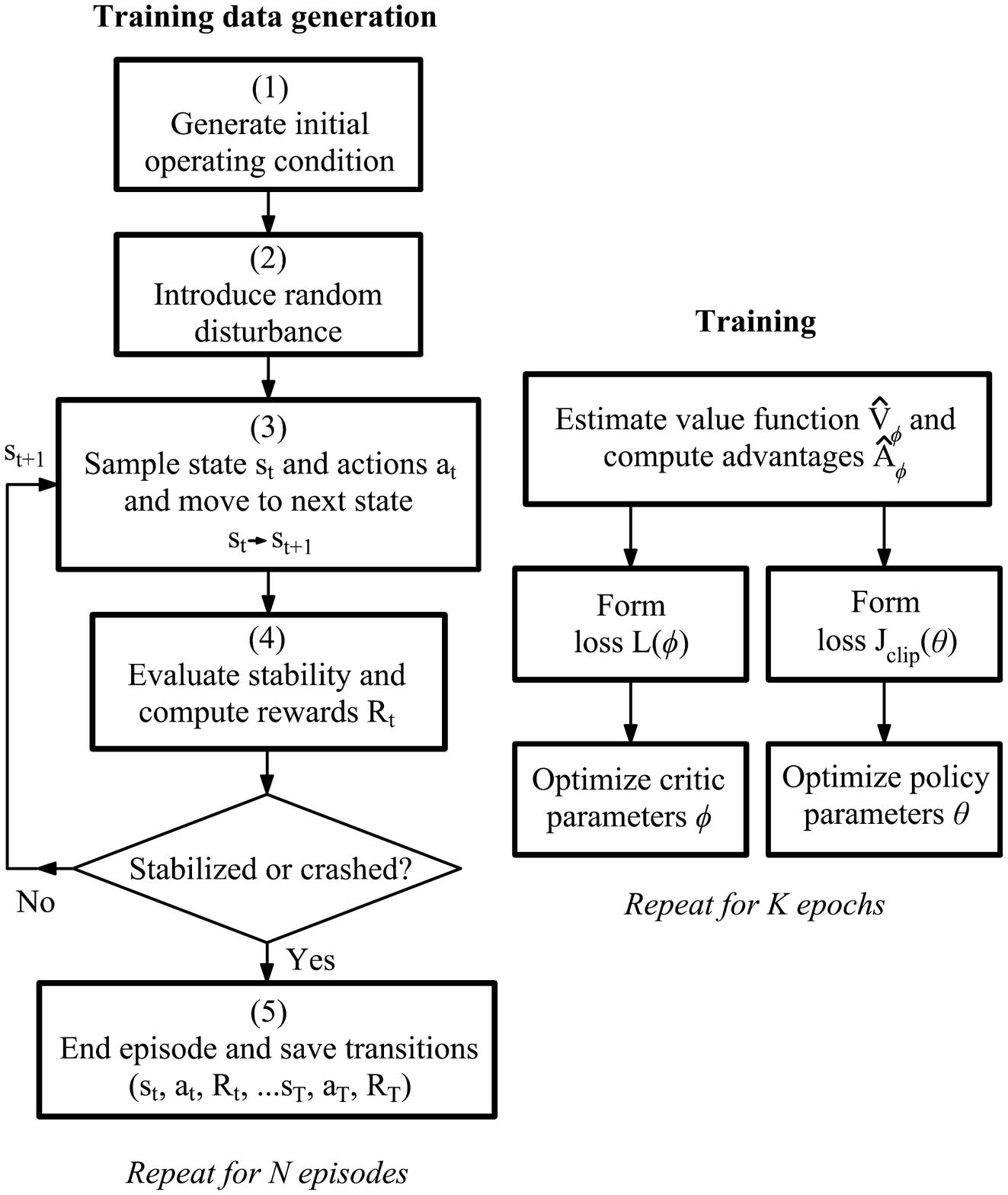}    
		\vspace{-0.1cm}
		\caption{Flowchart showing the generation of training data and the training of the actor and critic network.}
		\label{fig:Flowchart_DRL_LTVS}
		\vspace{-0.4cm}
	\end{center}
\end{figure}

An overview of the steps involved in generating training data and training the DRL agent is illustrated in Fig.~\ref{fig:Flowchart_DRL_LTVS}. The different steps are detailed in the sections below. 

\subsubsection{Generate initial operating condition}
For the Nordic32 test system, the initial operating conditions (OCs) were randomly generated around the insecure operation point denoted as "operating point A" in \cite{CutsemTestSystem2020}. All loads in the system were randomly and individually varied by multiplying the active load value with a random variable generated from a uniform distribution (95~\% of the original load as lower limit, 105~\% of the original load as upper limit). The power factor of all loads was kept constant. A load flow solution was then computed where any changes in total load in the system were compensated by the slack bus generator, \textit{g20}, see Fig.~\ref{fig:Nordic32_RL}. In actual implementations, the variations in load and generation should reflect the actual variation in that will occur in a given power system.

\subsubsection{Introduce random disturbance}
Once an initial OC was generated, a dynamic simulation was initialized and a single larger disturbance was introduced. The DRL agent was trained to handle different types of disturbances, and with the same probability for each scenario, either a line was tripped between buses (i) \textit{4032}-\textit{4044}, (ii) \textit{4032}-\textit{4042}, (iii) \textit{4031}-\textit{4041},  (iv) \textit{4021}-\textit{4042}, or the tripping of either (v) generator \textit{g6}, or (vi) generator \textit{g7}. The disturbances were chosen as they were proven to cause significant system stress in the "Central" area (see Fig.~\ref{fig:Nordic32_RL}), and without suitable control actions would in most load scenarios cause long-term voltage instability. In real applications, preferably all disturbances that are likely to cause long-term voltage instability should be evaluated and included in the training.  However, this would require significantly more training data to achieve satisfactory results and without a loss of generalization, we reduced the study to include only the previously mentioned disturbances. Furthermore, the DRL agent is in this study only trained on disturbance scenarios. In actual applications, it would be important to also include normal load scenarios in the training data to ensure that the DRL agent can learn to differentiate when it should activate emergency actions or not.

\subsubsection{Sample state $s_t$ and action $a_t$ from the policy and move to next state}

The state was then sampled from the system and passed to the actor network. The actor network outputs parameters that form the current policy $\pi_\theta(a|s)$ from which an action was sampled. That action was then activated in the system and the simulation continued to run until the next time step, which formed the state transition from $s_t \rightarrow s_{t+1}$. The time between each step in the simulation was 5 seconds (while the integration step size in the dynamic simulation in PSS\textsuperscript \textregistered E was 0.05 seconds). The states and actions are further discussed in section \ref{sec:states} and \ref{sec:actions}, respectively. 

\subsubsection{Evaluate stability and compute rewards $R_t$}
Once the dynamic simulation reached the next time step, the stability of the system was evaluated. If any transmission system bus voltage ($V_{TS}$) was below 0.7 pu at \textit{any} time in the dynamic simulation, the system was assumed to be unstable and the episode was terminated in advance. If any $V_{TS}$ at the \textit{end} of the dynamic simulation was below 0.90 pu, the system was \textit{also} assumed to be unstable. This allows for system voltages to decrease below 0.9 pu for shorter time periods but is still not acceptable in the long term. At every time step, a reward was also computed. The reward levels are designed to always incentivize the DRL agent to activate load curtailment rather than allowing the system to crash, while still minimizing action costs. The reward $R_t$ at each time step was a combination of the cost for the taken action ($C_{a}$);  a smaller penalty (-1) if any $V_{TS}$ were below 0.90 pu during the dynamic simulation; or a larger penalty (-500) if the system had become unstable at any point. The action cost $C_{a}$ is further discussed in Section \ref{sec:actions}. The reward at every time step was then computed as: 
\begin{equation} \label{eq:reward_computation}
    R_t= 
\begin{cases}
    C_{a} - (500\cdot0.99^t)   & \text{if unstable} \\
    C_{a}, & \text{else if all } V_{TS} \geq 0.90 \text{ pu} \\
    C_{a} - (1\cdot0.99^t) ,& \text{else if any } V_{TS} < 0.90 \text{ pu} \\
\end{cases}
\end{equation}
The penalties were multiplied by a discounting factor of $0.99^t$, resulting in lower negative penalties if instability and low system voltages occurred later rather than early in an episode. In this study, the reward is unitless, but should in real applications reflect the actual monetary cost of different actions and the corresponding rewards when the control goal is either achieved or missed.

\subsubsection{End episode and save transitions}
All episodes ran for a maximum of $T=1000$ seconds unless the system become unstable and was terminated in advance. At the end of all episodes, the transition data ($s_t$,$a_t$,$R_t$,...,$s_T$,$a_T$,$R_T$) was stored and later used during training.

\subsection{States}
\label{sec:states}
The states were sampled from measurements taken from the dynamic simulation and consisted of a vector of i) bus voltages magnitudes, ii) active power flows, and iii) reactive power flows of all (and in between) buses in the system. The states also included information about the current price of action activation and the capacity of load curtailment at each participating bus, further explained in section \ref{sec:actions}. While the relatively slow sample rate would allow measurements to be sampled from supervisory control and data acquisition (SCADA) systems, the availability and use of phasor measurement units would ensure a higher modeling accuracy through the time-synchronized measurements. To also capture the dynamics of the system, \textit{previous} observations from time step $t$-$1$ were stacked and included in the state vector (thus doubling the length of the state vector). The state vectors were then normalized by subtracting the mean value of each state value by its mean and then dividing by the standard deviation. The mean and standard deviation of each state value was computed from previously sampled states and a list with a maximum of 10~000 sampled states was stored. Once 10~000 sampled states were added to the list, the mean and standard deviation used for normalizing states became fixed. 

\subsection{Actions}
\label{sec:actions}
To stabilize the system in case of instability, the DRL agent could activate load curtailment resources that, for instance, ESS and DR could provide to system operators. ESS and DR can essentially be viewed as available load curtailment that has been procured through a market system \cite{Rahimi2010}. Using such services would allow system operators to alleviate stress in a power system similar to that of load shedding. The difference is mainly i) that a higher degree of flexibility is available, where the activation level of the load curtailment based on ESS/DR can typically be taken in much smaller steps than load shedding, and ii) that the impact on the end-users would be significantly less than when using forced load shedding. 

The DR and ESS are modeled implicitly by allowing the DRL agent to adjust the load levels at participating load buses within a certain range. At each time step, the DRL agent takes a single continuous action. The continuous action is used to control the total load curtailment in the system, which is then distributed among participating load buses. For simplicity and to allow clearer visualization of the results, only two load buses are participating in the load curtailment, which are located at bus 1044 and bus 1045. The availability and the price of market-based system services provided by DR and ESS are typically varying; an uncertainty that needs to be included in the training of the DRL agent. To model this, the level of load curtailment that is available at each of the two participating buses is varied at the beginning of each disturbance scenario. The capacity of load curtailment is determined by sampling from a random uniform distribution with a lower level of 300~MW, and an upper level of 500~MW. Furthermore, the price of activating the load curtailment is also varied between the two participating buses which is achieved by randomly varying the price for each bus at the beginning of each disturbance scenario. The price of activating load curtailment at each bus is also determined by sampling from a random uniform distribution with a lower cost of -0.1/MW, and an upper level of -0.2/MW. 

The DRL agent then controls the \textit{total} level of load curtailment that is taken at each time step. The bus with the lowest price is activated first, but if it has not sufficient capacity in adjusting its load, the other bus (with a higher price) will be activated as well. The range of load curtailment capacity was chosen to ensure that all of the load and disturbance scenarios could be stabilized if sufficient load curtailment was utilized, while still adding uncertainty in where the load curtailment was activated. The price variation models a market-based system, where the price of ESS/DR will typically vary depending on availability and the types of load/customers and storages that are connected to the service. Thus, the chosen approach ensures that there will be an uncertainty in \textit{where} the actions are activated, to \textit{what level} the actions are available, and also to \textit{what cost} to the system. Depending on which bus the load curtailment is activated on may also impact the effectiveness of the control action to mitigate instability, which is an additional uncertainty that the DRL agent needs to account for.

In this study, the availability of the modeled ESS/DR was chosen to always be sufficient to stabilize the system. While only the case of two participating buses was analyzed, the method could be trained to have providers of ESS/DR at \textit{any} bus. Furthermore, in actual applications, if the total required load curtailment is higher than what could be supported by participating ESS/DR systems, conventional load curtailment in the form of load shedding may be required as a complement. In such cases a combination of ESS/DR and load shedding could be used, where load curtailment by ESS/DR is always activated first to minimize the level of required load shedding.

\subsection{Architecture of actor and critic network}
\label{sec:network}
\begin{figure}
	\begin{center}\vspace{-3.0cm}		\includegraphics[width=8.2cm]{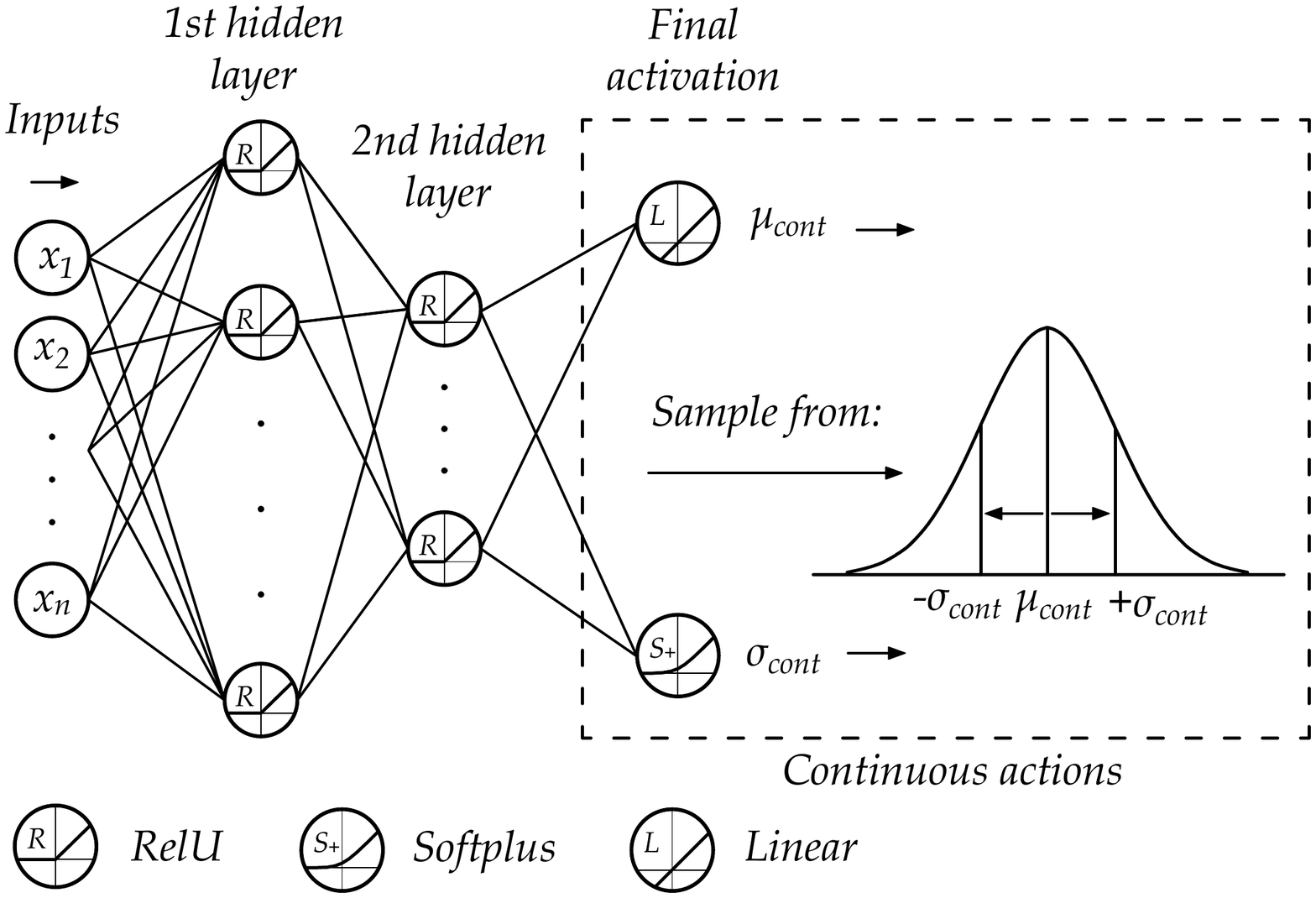}    
		\vspace{-3.3cm}
		\caption{Architecture of the actor network.}
		\label{fig:RL_Architecture_Tight_LTVS}
		\vspace{-0.5cm}
	\end{center}
\end{figure}

The actor network, illustrated in Fig.~\ref{fig:RL_Architecture_Tight_LTVS} and further detailed in Table \ref{table:NN_Architectures}, forms the mapping from states to the stochastic policy $\pi_\theta(a|s)$, from which actions are sampled. The network has two hidden layers followed by a final activation layer with two different activation functions used to form the outputs. The network outputs parameters used in defining a Normal distribution $\mathcal{N}$ from which the policy is defined and actions are sampled:
\begin{equation} \label{eq:Normal_Dist}
\pi_\theta(a|s) = \mathcal{N} \left(\mu_{\theta}(s),\sigma^2_{\theta}(s)\right)
\end{equation}
The Normal distribution is parametrized by a mean value $\mu_{\theta}$ and a standard deviation $\sigma_{\theta}$, where the mean value $\mu_{\theta}$ is computed using a linear activation function in the final layer, while the standard deviation $\sigma_{\theta}$ is computed using a softplus activation function that ensures that the value never becomes negative. The stochastic formulation of the policy is used to allow exploration of the action space. The critic network is separate from the actor network and consists of a simple NN with a single hidden layer and a linear final activation function, further detailed in Table \ref{table:NN_Architectures}. 

\subsection{Training of actor and critic networks}
Once a total of $N=64$ episodes were sampled, the actor and the critic networks were trained. The critic network was first used to compute the value $\hat{V}_\phi^\pi$ of each state. Once the value of each state was computed, the estimated advantage of each state was computed using (\ref{eq:est_AdvFunc}). The cost function used to train the actor network is computed using (\ref{eq:PPO_clipped}) on all samples for all $N$ episodes. Once the cost function for all samples was computed, the final $J^{clip}(\theta)$ was computed by taking the mean of those values. The cost function used to train the critic network, $L(\phi)$, is computed by taking the mean squared error on all $\delta$-errors from (\ref{eq:est_AdvFunc}) for all samples and all $N$ episodes, followed by computing the mean of those values. The training was performed using the software Tensorflow in Python which automatically computes the gradients on the defined cost functions. The training was performed for $K=5$ epochs on the whole batch of $N$ episodes simultaneously. The values of the learning rates and other hyperparameters used in the training are specified in Table \ref{table:NN_Architectures}. Once the networks were trained on the stored transition data, new training data were generated and the old data that were sampled with the old policy were discarded.

\begin{table}
	\caption{Design and hyperparameters used in training.}
	\label{table:NN_Architectures}
	\vspace{-0.3cm}
	\centering
	\begin{tabular}{l l l l}
		\toprule
		& & \textbf{Parameter} & \textbf{Values}\\ 
		\midrule
						\addlinespace
		\multirow{10}{*}{{\rotatebox[origin=c]{90}{\textit{Architecture}}}} & \multirow{4}{*}{{\rotatebox[origin=c]{90}{\textit{Critic}}}}
		& Number of inputs &  976 \\
		& & Neurons in hidden layer & 128  \\
		& & Final activation function & Linear \\
		& & Hidden layer activation & RelU \\
		\addlinespace
		\addlinespace
		 & \multirow{6}{*}{{\rotatebox[origin=c]{90}{\textit{Actor}}}} &
	    Number of inputs &  976 \\
		& & Neurons in 1st hidden layer & 64  \\
		& & Neurons in 2nd hidden layer & 32  \\
		& & Final activation for $\mu_{cont}$ &  Linear \\
		& & Final activation for $\sigma_{cont}$ &  Softplus \\
		& & Hidden layer activation & RelU \\
		\addlinespace
		\addlinespace
		 & \multirow{6}{*}{{\rotatebox[origin=c]{90}{\textit{Training}}}}  &
		Epochs ($K$) &  5 \\
		& & PPO clip parameter ($\epsilon$) &  0.2 \\
		& & Optimizer & Adam \cite{Kingma2014} \\
		& & Batch size ($N$) & 64  \\
		& & Actor network learning rate & 5e-4  \\
		& & Critic network learning rate & 5e-3  \\
		
				\addlinespace
		\bottomrule
		\vspace{-0.7cm}
	\end{tabular}		
\end{table}

\section{Simulations and results}
\label{sec:Simulations}

\subsection{Training results}

\begin{figure}
	\begin{center}\vspace{-0.1cm}		\includegraphics[width=9.4cm]{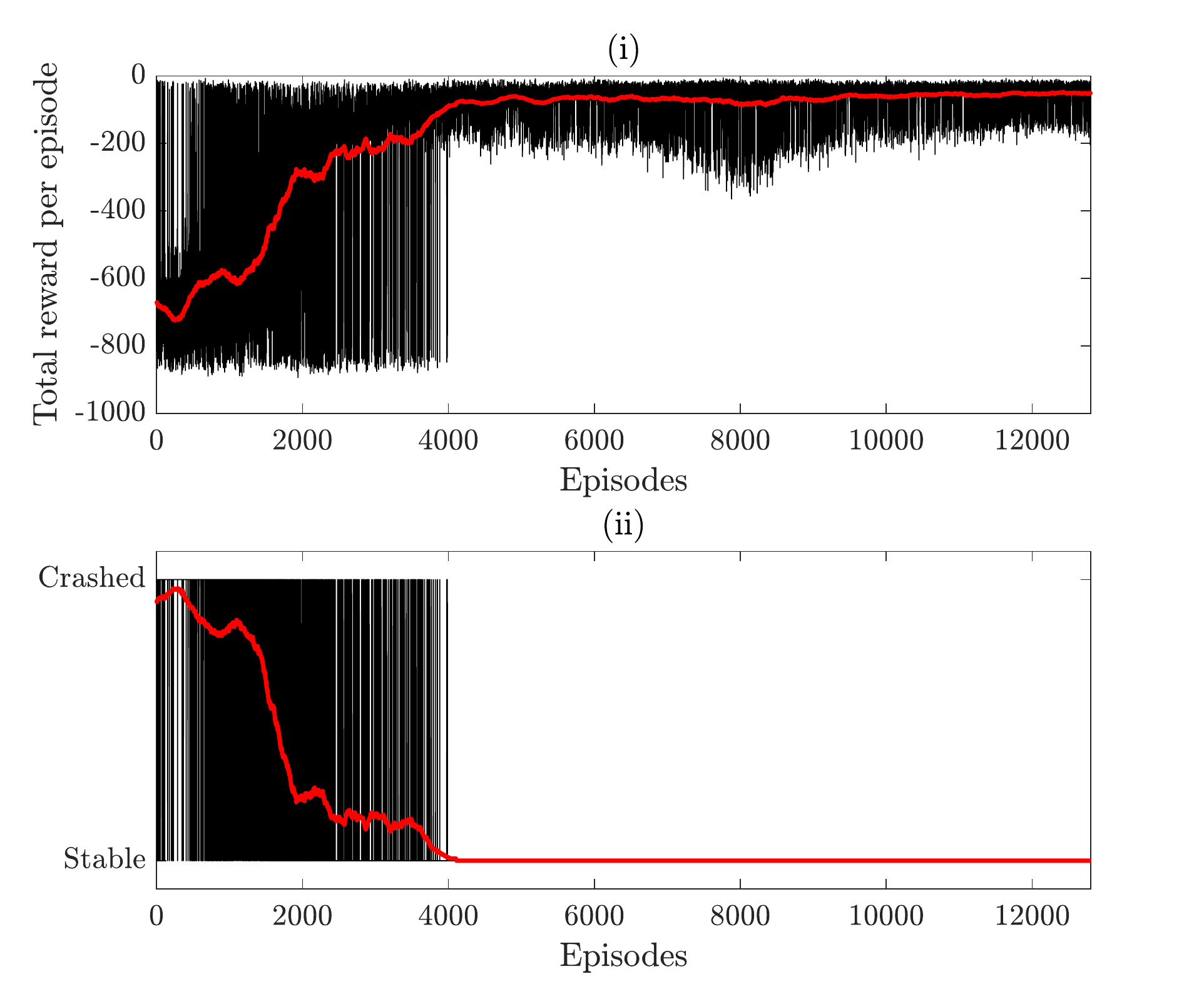}    
		\vspace{-0.9cm}
		\caption{Performance and development over episodes: Sub-figures showing (i) total reward per episode; (ii) whether the episode resulted in a crash or a stable state at the end of the simulation. The red line indicates a moving average computed over the mean of 250 data points.}
		\label{fig:Training_results_LTVS_control}
		\vspace{-0.4cm}
	\end{center}
\end{figure}

The DRL agent was trained for a total of 12 800 episodes, after which the performance converged. The training performance is shown in Fig.~ \ref{fig:Training_results_LTVS_control}, where the total rewards per episode and whether the episode resulted in a crash or a stable state at the end of the simulation, is shown. The red line shows a centered moving average computed over the mean value over 250 episodes. The results in sub-figure (i) show that the performance improved rapidly until around 4~000 episodes, after which the policy managed to avoid system collapses completely. After this, the performance continued improving by mainly optimizing the level of action activation for each of the scenarios. 

\subsection{Test sets}
\label{sec:test_sets}

During training, the DRL agent used a stochastic policy which allowed it to automatically explore the available action space. However, when using it online it is more suitable to transform the policy into a deterministic one and always pick the actions that with the \textit{highest probability} are optimal. When testing the algorithm, the continuous action was thus not sampled from a normal distribution, but rather controlled directly by the mean value $\mu_{cont}$ which was one of the outputs of the actor network. The trained DRL agent was tested on three different test sets. A total of 100 test scenarios were computed for each test set. The test sets used were defined as: 
\begin{enumerate}
    \item \textbf{Test set 1}: Data generated in the same way as for the training data, but using a deterministic policy instead.
    \item \textbf{Test set 2}: Introducing new unseen OCs by increasing the variation of the generation and load configurations. Instead of randomly adjusting each load between 95~\% to 105~\% as specified in Section \ref{sec:GenerationOCs}, the OCs were adjusted randomly between 90~\% to 110~\%. 
    \item \textbf{Test set 3}: Introducing new unseen OCs by introducing a disturbance that was not used in training the DRL agent. The new disturbance is the tripping of the line between the buses \textit{4011}-\textit{4021}. The same variation of generation and load configuration as during training was used. 
\end{enumerate}

The performance of the developed DRL agent was also compared to that of a rule-based load shedding scheme. The load shedding scheme acts whenever \textit{any} transmission system voltage is below 0.90 pu; a threshold value which is in the range found suitable for the Nordic32 test system in \cite{Otomega2007}. In that case, a total of 100 MW load is removed from the system, divided equally between the loads located at bus 1044 and 1045. The load shedding scheme could be activate at any time step throughout the whole simulation time. The 5 seconds interval between each time step prevented the load shedding scheme from reacting to voltage dips during the short transient period just after each disturbance. To allow a fair comparison, the cost for activating the load shedding ($C_a$) was chosen to \text{-0.15/MW}, which is the mean value of the varying price for activation of the DR/ESS resources used by the DRL agent. The main purpose of the comparison is to illustrate the strengths and weaknesses of the DRL control when compared to a more conventional load shedding scheme. Further enhancement to the rule-based load shedding scheme could be achieved by more intricate settings for when and how much load is being disconnected.

\subsection{Test performance}
The average reward on the different test sets is presented in Table~\ref{table:Average_performance_test} and is computed as the mean total reward per episode for all test scenarios. In the final column, the relative difference between the DRL control and the load shedding scheme is presented. The results show that the DRL agent managed to get a significantly lower negative average reward compared to the load shedding control scheme on \textit{all} different test sets. For instance, in test set 1, the load shedding scheme resulted in a 128.6~\% higher \textit{negative} reward compared to when the DRL control was used. Although not explicitly being trained on the load and disturbance scenarios found in test set 2 and test set 3, the DRL agent managed to generalize its learning to these scenarios and still find a significantly more efficient control policy than for the load shedding scheme. The improvement was smallest on test set 3 (33.5~\%) when a new disturbance that was not included in the training data was used to stress the system. It should be noted that all test scenarios in each test set were successfully controlled to stable states, both for the DRL control and the rule-based load shedding scheme. 

\begin{figure}[t]
	\begin{center}\vspace{-0.2cm}		
	\includegraphics[width=8.4cm]{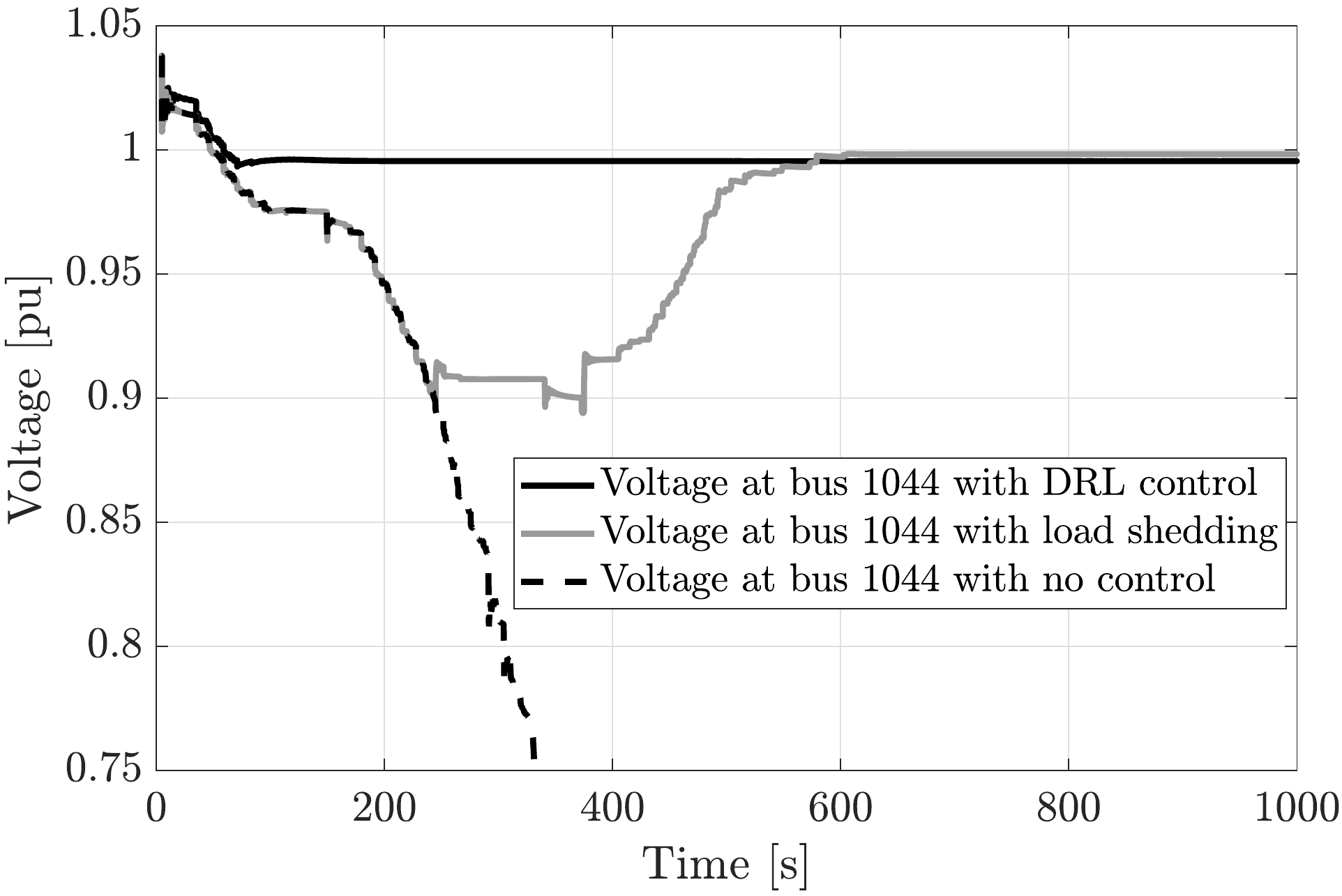}    
		\vspace{-0.0cm}
		\caption{Voltage magnitude at bus 1041 over time for different control schemes. }
		\label{fig:Volts1041}
		\vspace{-0.35cm}
	\end{center}
\end{figure}

\begin{figure}[t]
	\begin{center}\vspace{-0.1cm}	
	\includegraphics[width=8.4cm]{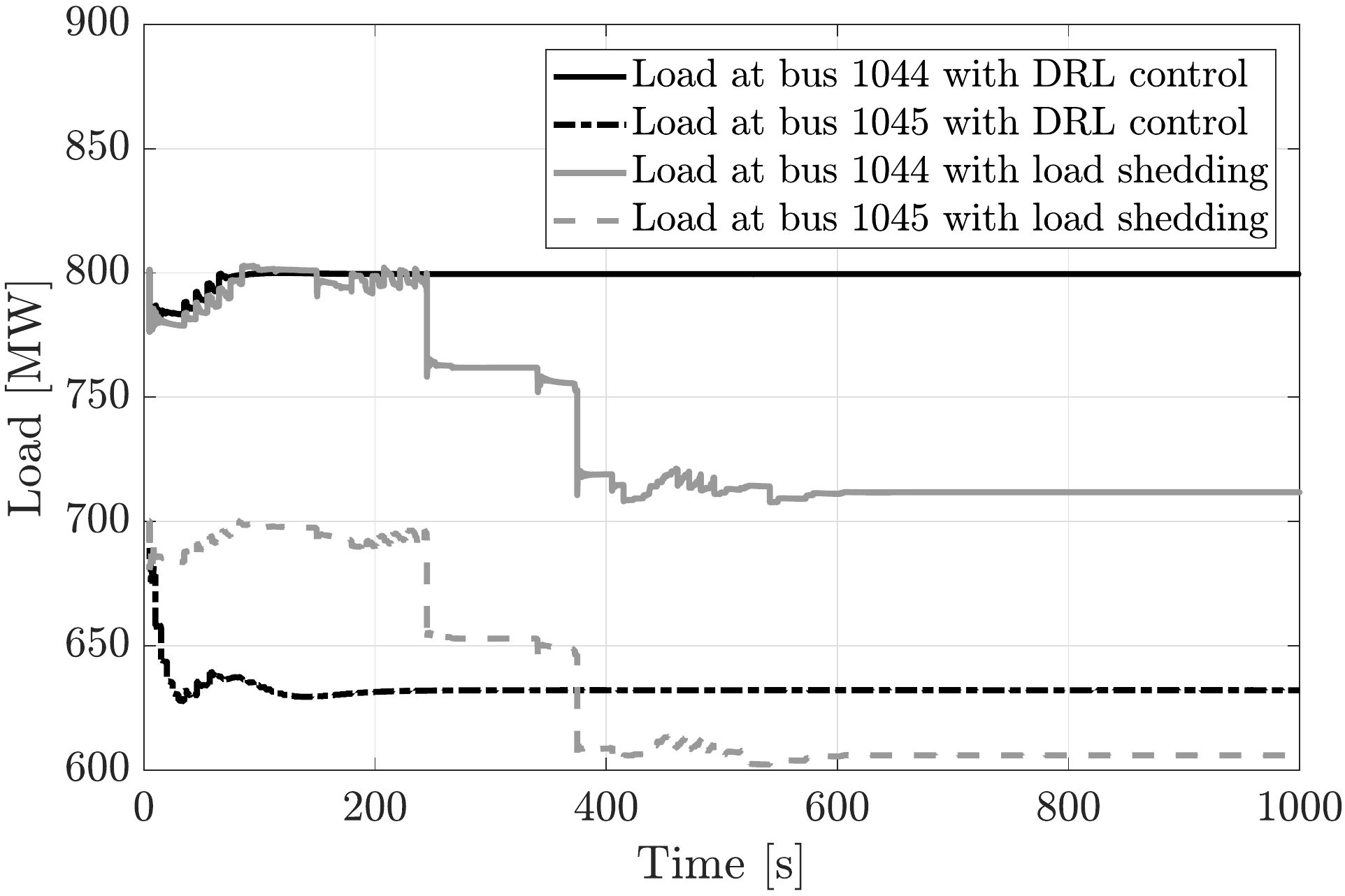}    
		\vspace{-0.0cm}
		\caption{Load development at bus 1044 and bus 1045 for the developed DRL control and for a load shedding control scheme. }
		\label{fig:Loads}
		\vspace{-0.35cm}
	\end{center}
\end{figure}

In Table~\ref{table:Average_load}, the average required load curtailment for each test set and control method is presented. This metric represents how much load each control method required to be curtailed before the system was stabilized. Once again, the relative difference between the two control methods is presented in the final column in the table. The results show that the DRL agent required significantly less load curtailment to stabilize the system compared to the load shedding scheme for all of the test sets. For instance, for test set 2, the load shedding required 259.4~\% more load in average to be curtailed compared to what was used by the DRL control. The differences between the DRL control and the control that is achieved with load shedding are exemplified in Fig.~\ref{fig:Volts1041} and Fig.~\ref{fig:Loads}. In Fig.~\ref{fig:Volts1041} the voltage magnitude at bus 1041 is shown for one of the test scenarios in test set 1. The voltage magnitude over time is presented for the cases i) when the DRL control is used, ii) when the load shedding control is used, iii) and when no control is used. In Fig.~\ref{fig:Loads}, the load at the controlled load buses 1044 and 1045 are also shown, which shows the difference in how the load is controlled by the DRL control and when using a load shedding control.

\begin{table}
	\caption{Average performance required for different test sets and control methods.}
	\label{table:Average_performance_test}
	\vspace{-0.2cm}
	\centering
	\begin{tabular}{lcccc}
		\toprule
		& \multicolumn{2}{c}{Mean total reward per episode} & Difference \\
		&  \textit{DRL control} & \textit{Load shedding} & [\%]  \\
		\midrule
		\addlinespace
		Test set 1 & -37.9  & -86.7 &  128.6~\% \\
		Test set 2 & -36.5  & -114.2  & 213.9~\%\\
	    Test set 3 & -16.7  & -22.3  & 33.5~\%\\
		\bottomrule
		\vspace{-0.7cm}
	\end{tabular}		
\end{table}

\begin{table}
	\caption{Average load curtailment required for different test sets and control methods.}
	\label{table:Average_load}
	\vspace{-0.2cm}
	\centering
	\begin{tabular}{lcccc}
		\toprule
		& \multicolumn{2}{c}{Average load curtailment [MW]} & Difference \\
		&  \textit{DRL control} & \textit{Load shedding} & [\%]  \\
		\midrule
		\addlinespace
		Test set 1 & 190.9  & 560.0 & 193.4~\% \\
		Test set 2 & 166.1  & 597.0  & 259.4~\% \\
	    Test set 3 & 124.5  & 144.0  & 15.7~\% \\
		\bottomrule
		\vspace{-0.7cm}
	\end{tabular}		
\end{table}

For the given scenario, the system will collapse after around 330 seconds if no control is initiated. For the case with load shedding, a total of 200 MW is shed from the system. The load shedding is activated once at around 250 seconds, and then another activation occurs at around 380 seconds, which can be seen from the relatively large steps in load reduction in Fig.~\ref{fig:Loads}. After the second activation of the load shedding, the system voltages are restored in the system, which can be seen in Fig.~\ref{fig:Volts1041}. For the DRL control, the load curtailment is activated \textit{directly} after the disturbance and in smaller increments, with no need to wait for the system to degrade before the control is activated. The load at bus 1045 is reduced by approximately 90 MW, after which the system is stabilized. The DRL control also manages to achieve a more satisfactory post-disturbance voltage magnitude profile, where the voltage magnitude is kept closer to the nominal pre-disturbance level. Thus, although the DRL control required a smaller amount of load curtailment, it achieved both a faster and more efficient control for the given scenario. Since the DRL control acts \textit{before} the system voltages have deteriorated, a lower level of load curtailment was required. The smoother control that is possible when utilizing load curtailment resources from DR and/or ESS also provided a more efficient way to mitigate voltage instability to a low system cost. 

\subsection{Performance with an action activation threshold}

One of the advantages of the DRL control is that \textit{any} level of load curtailment can be activated at each time step, while typically a load shedding scheme is activated in significantly larger blocks of load curtailment. However, while the DRL agent is trained to minimize the control actions once the system has stabilized, it is difficult to train the action (controlled by the mean value $\mu_{cont}$) to \textit{fully} converge to zero. Furthermore, when evaluating the DRL agent's performance on stable disturbance scenarios (i.e. disturbance scenarios that would end up being stable despite \textit{no} load curtailment being activated), it was found that the DRL agent did (unnecessarily) activate a small level of load curtailment. The reasons for this behavior can mainly be explained by the fact that the DRL agent was trained on a majority of cases that were not secure, which can be observed by noting the number of crashed scenarios at the beginning of the training by the red line in sub-figure (ii) in Fig. \ref{fig:Training_results_LTVS_control}.

Avoiding unnecessary activation of load curtailment will be important if the DRL control is to be included in any real control systems. The main solution would be to train the DRL agent on more stable scenarios to make it more robust in handling such scenarios. This could however be combined with an action activation threshold, to make sure that only relatively significant action signals are activated in the system. To test this feature, an activation threshold of $\pm$10 MW of load curtailment was applied for the DRL control. Thus, any control action from the DRL agent with a lower magnitude than 10~MW resulted in \textit{no activation} of load curtailment, while any control action larger or equal to 10~MW was activated. 

The action activation threshold was tested on the same test sets that were developed in Section \ref{sec:test_sets}. The results are presented in Table \ref{table:Average_performance_actionlimit}, where the mean total reward per episode and the average required load curtailment are presented when the action activation threshold was activated. The percentage difference in performance compared to the case when no action activation threshold was implemented (while still using the DRL control) is presented in parenthesis after each value. The level of average load curtailment reduced significantly compared to the the case when no action activation threshold was implemented, for all of the sets. The difference was most significant for test set 3, where the average load curtailment was reduced by \text{-82.1~\%}. However, the mean total reward per episode worsened for test 1 and test 2, where the (negative) mean total reward per episode increased by 32.7~\% and 13.2~\%, respectively. The higher negative rewards were caused by the transmission system voltages remaining below 0.9 pu for a longer time during the post-disturbance state, which resulted in a higher negative reward for those scenarios.

\begin{table}
	\caption{Average performance and load curtailment when using an action activation threshold.}
	\label{table:Average_performance_actionlimit}
	\vspace{-0.2cm}
	\centering
	\begin{tabular}{lcc}
		\toprule
		& Mean total  & Average load curtailment \\
		& reward per episode & [MW]  \\
		\midrule
		\addlinespace
		Test set 1 & -50.3 (32.7~\%)  & 171.6 (-10.1~\%) \\
		Test set 2 & -41.3 (13.2~\%) & 114.2 (-29.1~\%) \\
	    Test set 3 & -12.4 (-25.8~\%) & 22.3 (-82.1~\%) \\
		\bottomrule
		\vspace{-0.4cm}
	\end{tabular}		
\end{table}

\begin{figure}
	\begin{center}\vspace{-0.0cm}		
	\includegraphics[width=8.6cm]{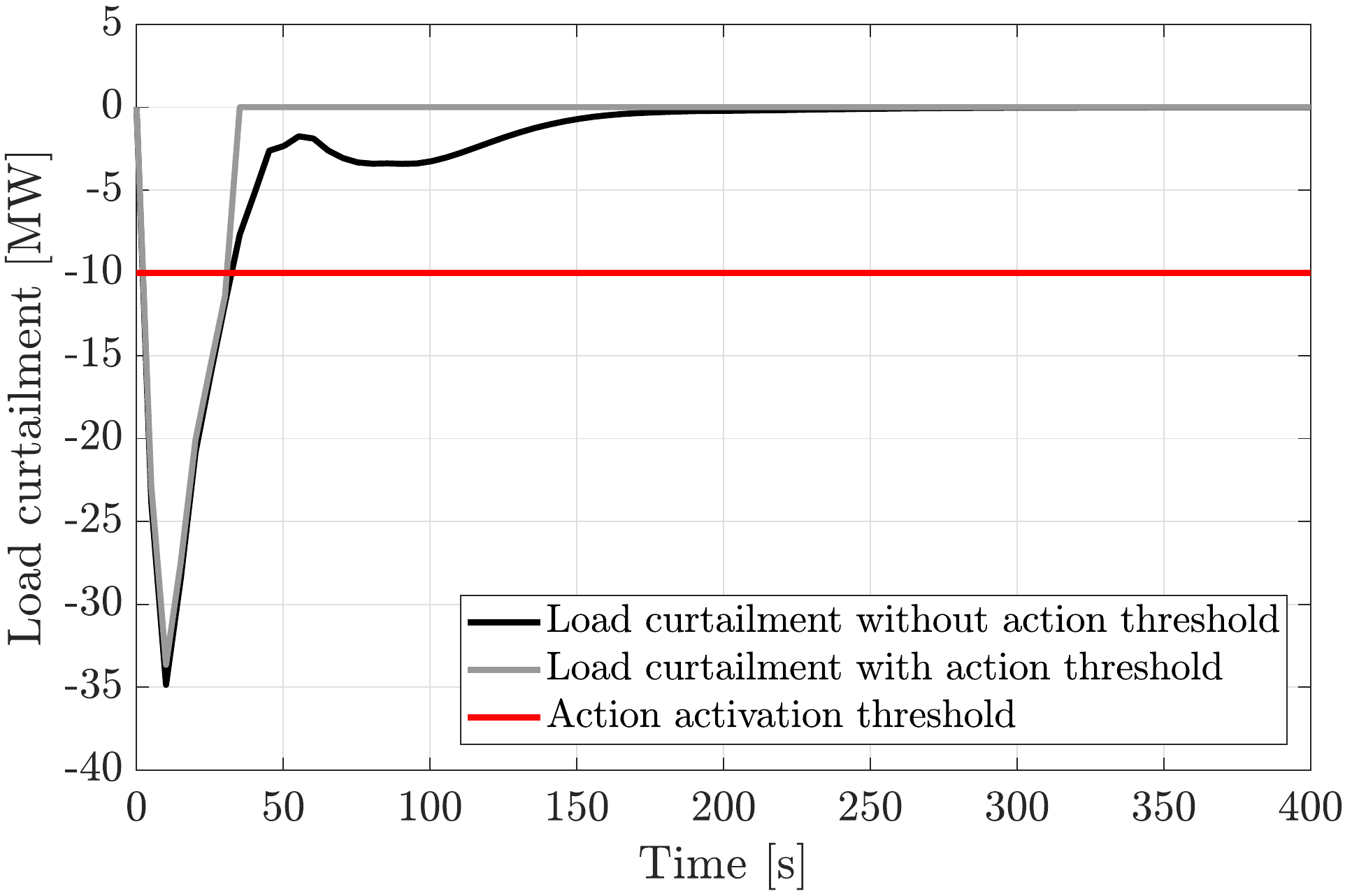}    
		\vspace{-0.2cm}
		\caption{Load curtailment activated for the DRL control with and without action activation threshold implemented. Time axis limited for better visualization or results.}
		\label{fig:Load_curtailment_limited}
		\vspace{-0.1cm}
	\end{center}
\end{figure}

\begin{figure}
	\begin{center}\vspace{-0.2cm}		
	\includegraphics[width=8.6cm]{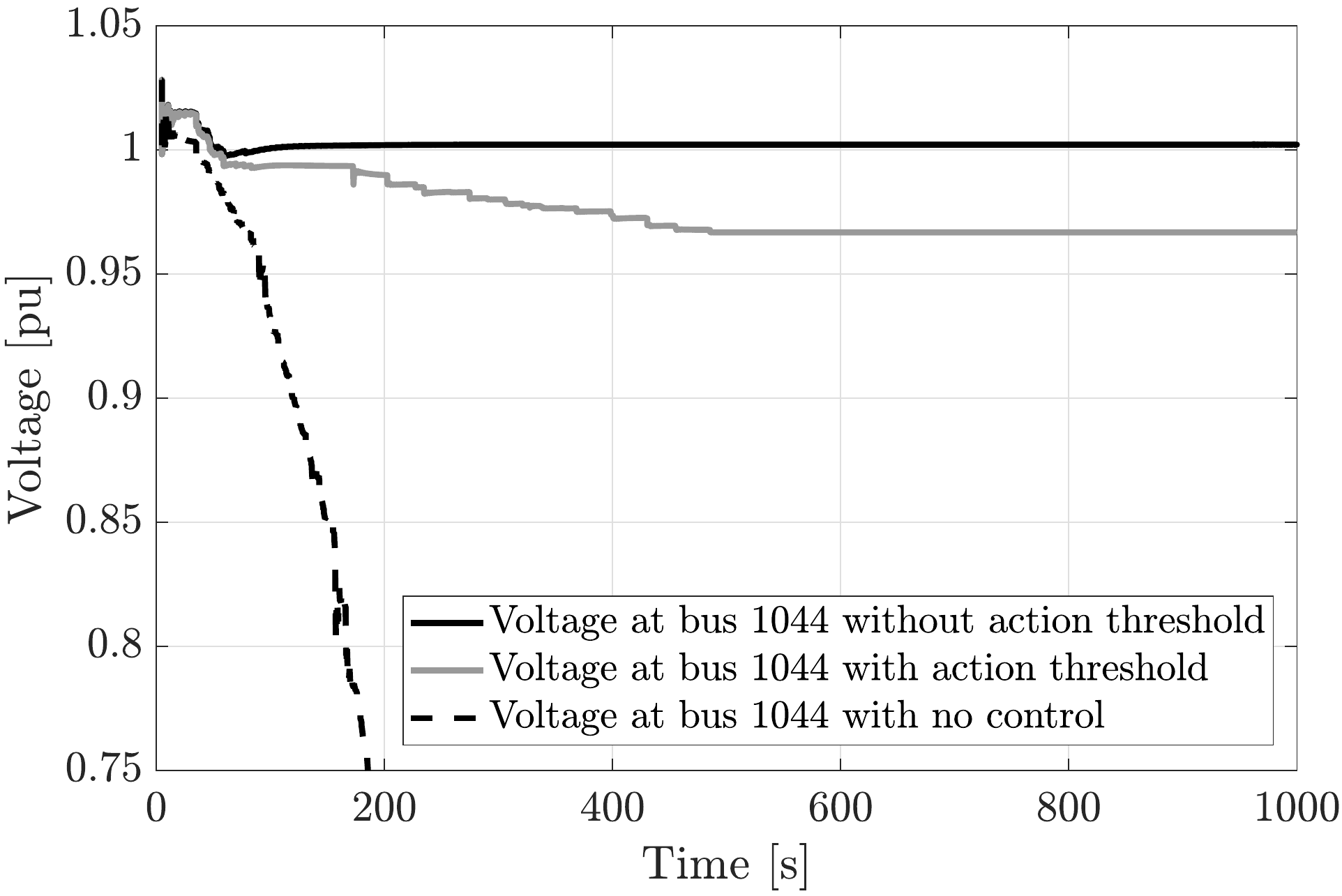}    
		\vspace{-0.2cm}
		\caption{Voltage magnitude at bus 1041 over time with or without an action activation threshold.}
		\label{fig:Volts1041_limited}
		\vspace{-0.1cm}
	\end{center}
\end{figure}

In Fig. \ref{fig:Load_curtailment_limited}, the action activation threshold is exemplified for a test scenario when both the action activation was used and when it was turned off. When the action activation threshold was used, the load curtailment was activated only in a short period after the disturbance, while in the case when no threshold was used, the load curtailment continued (albeit with low low levels of activation) up to around 200 seconds after the disturbance occurred in the system. In Fig. \ref{fig:Volts1041_limited}, the resulting voltage magnitude at bus 1041 is shown for i) the case with no action threshold, ii) with the action threshold used, and iii) when no control is used for the same test scenario. The results show a significantly lower post-disturbance voltage magnitude over time when the action activation threshold was used, but the system still managed to remain stable.

\section{Conclusions}
\label{sec:Conclusions}

This paper introduces a control method based on DRL to mitigate long-term voltage instability events in real-time. Once trained, the DRL control can continuously assess the system stability and suggest fast and efficient control actions to system operators. The DRL control is trained to use system services from DR and ESS as a more efficient and flexible alternative to stabilize the system compared to load shedding. The uncertainty in availability and the price of such market-based system services is modeled in the system. The developed DRL control was tested on a modified version of the Nordic32 test system and showed good performance on all of the developed test sets. The DRL control was also compared to a more conventional rule-based load shedding scheme and was shown to provide a more efficient and fast control. An action activation threshold was implemented to reduce load curtailment activation once the system was already stabilized. 

Future research work includes: i) extending the study to include more actions spaces; ii) adapt and expand the control to also handle other types of stability phenomena iii) further evaluating the generalization capability of DRL control to handle scenarios not included in the training; iv) evaluating the impact of a delay between actions taken by the DRL agent and them being activated in the system; v) evaluating recent advancements in safe DRL to address control challenges in safety-critical systems such as power systems.

\section*{Acknowledgment}
The work presented in this paper has been financially supported by Energiforsk
and Svenska kraftnät (Swedish National Grid) under project numbers
EVU10140 and EVU10450.

\IEEEtriggeratref{8}

\bibliography{Referenser}
\label{sec:Referenser}

\end{document}